\documentclass[fleqn,usenatbib]{mnras}
\usepackage[utf8]{inputenc}

\usepackage[T1]{fontenc}
\usepackage{ae,aecompl}
\usepackage{newtxtext,newtxmath}

\usepackage{graphicx}
\usepackage{amsmath}
\usepackage{color}
\usepackage{hyperref}
\hypersetup{colorlinks=true, linkcolor=blue, filecolor=magenta, urlcolor=cyan}

\newcommand{\be}{\begin{equation}}
\newcommand{\ee}{\end{equation}}
\newcommand{\bea}{\begin{eqnarray}}
\newcommand{\eea}{\end{eqnarray}}
\newcommand{\omgw}{\Omega_\mathrm{GW}}
\newcommand{\msol}{\mathrm{M}_\odot}

\title[Astrophysical gravitational-wave background]{Astrophysical Uncertainties in the Gravitational-Wave Background from Stellar-Mass Compact Binary Mergers}

\author[L. Lehoucq et al.]{Léonard Lehoucq$^{1}${\thanks{Email: lehoucq@iap.fr}}, Irina Dvorkin$^{1}$, Rahul Srinivasan$^{2,3}$, Clément Pellouin$^{1}$, Astrid Lamberts$^{2,3}$\\
$^{1}$Institut d'Astrophysique de Paris, Sorbonne Université and CNRS, UMR 7095, 98 bis bd Arago, F-75014 Paris, France\\
$^{2}$Université Côte d’Azur, Observatoire de la Côte d’Azur, CNRS, Lagrange, France\\
$^{3}$Université Côte d’Azur, Observatoire de la Côte d’Azur, CNRS, ARTEMIS, France\\}

\date{Accepted XXX. Received YYY; in original form ZZZ}

\pubyear{2023}
\volume{XXX}
\pagerange{1--12}

\begin{document}

\maketitle

\begin{abstract}

\noindent We investigate the Stochastic Gravitational Wave Background (SGWB) produced by merging binary black holes (BBHs) and binary neutron stars (BNSs) in the frequency ranges of LIGO/Virgo/Kagra and LISA. We develop three analytical models, that are calibrated to the measured local merger rates, and complement them with three population synthesis models based on the COSMIC code. We discuss the uncertainties, focusing on the impact of the BBH mass distribution, the effect of the metallicity of the progenitor stars and the time delay distribution between star formation and compact binary merger. We also explore the effect of uncertainties in binary stellar evolution on the background. For BBHs, our analytical models predict $\omgw$ in the range $[4\cdot10^{-10}-~1\cdot10^{-9}]$ (25~Hz) and $[1\cdot10^{-12}-~4\cdot10^{-12}]$ (3~mHz), and between $[2\cdot10^{-10}-~2\cdot10^{-9}]$ (25~Hz) and $[7\cdot10^{-13}-~7\cdot10^{-12}]$ (3~mHz) for our population synthesis models. This background is unlikely to be detected during the LIGO/Virgo/Kagra O4 run, but could be detectable with LISA. We predict about 10 BBH and no BNS mergers that could be individually detectable by LISA for a period of observation of 4 years. Our study provides new insights into the population of compact binaries and the main sources of uncertainty in the astrophysical SGWB.

\end{abstract}

\begin{keywords}
black hole mergers -- neutron star mergers -- gravitational waves
\end{keywords}

\section{Introduction}\label{sec:intro}

Gravitational waves (GW) have emerged as a promising tool for exploring our Universe \citep{PhysRevLett.116.061102}.
In recent years, ground-based detectors such as the Laser Interferometer Gravitational-Wave Observatory (LIGO) and Virgo have detected GWs from compact binary (CB) mergers. To date, a total number of $90$ CB mergers with an inferred probability of astrophysical origin of $p_\mathrm{astro} > 0.5$ have been detected \citep{Abbott_2019,Abbott_2021,theligoscientificcollaboration2022gwtc21,theligoscientificcollaboration2021gwtc3,the_ligo_scientific_collaboration_population_2022}. 
These observations have already provided valuable information about black hole (BH) and neutron star (NS) populations, in particular they have permitted to refine stellar population models that predict the formation and merger rates of CBs \citep[e.g.][]{2021ApJ...915L..35K,2021ApJ...910..152Z,Srinivasan+23,mapelli_properties_2019,Baibhav2019,2022MNRAS.516.5737B,van_Son_2022}.

In addition to these resolved signals from individual binary mergers, the incoherent superposition of unresolved sources creates an astrophysical Stochastic Gravitational Wave Background (SGWB). This signal, if detected, contains crucial information about high-redshift CB mergers. Moreover, other astrophysical and cosmological sources could contribute to the SGWB, including core-collapse supernovae, rotating neutron stars, primordial BHs, cosmological inflation, cosmic strings and first order phase transitions in the early Universe \citep[see][for extensive reviews]{2011RAA....11..369R,2018CQGra..35p3001C,christensen_stochastic_2019,2022Galax..10...34R}.

The Laser Interferometer Space Antenna (LISA), a space-based interferometer to be launched in the next decade, aims to detect new classes of GW sources at mHz frequencies. One of the primary objectives of LISA is to detect a cosmological SGWB, which could, for example, be produced by phase transitions in the primordial Universe or by cosmic strings \citep{2018CQGra..35p3001C,auclair2022cosmology}. Most cosmological backgrounds are expected to be significantly lower in magnitude than the astrophysical one, however some realistic cases rooted in particle physics, such as cosmic strings, predict signals which can be orders of magnitude larger \citep{2020JCAP...04..034A}. Nonetheless, investigating the properties of the astrophysical background is crucial in order to prepare the detection strategies of the cosmological signal \citep{cusin_stochastic_2020,chen_stochastic_2019,zhao_stochastic_2020,liang_science_2022}.

The investigation of the SGWB is an active field of research, both theoretically and observationally. Data collected by the LIGO and Virgo observatories during the first three observational runs have been used to search for this background in the $\mathcal{O}$(10)-$\mathcal{O}$(100)~Hz range, but so far only upper limits have been inferred \citep{PhysRevD.104.022005,PhysRevD.104.022004,PhysRevD.100.061101,PhysRevLett.120.201102,PhysRevLett.120.091101,PhysRevLett.119.029901}. The most stringent upper limit provided by the LIGO/Virgo collaboration in \citet{PhysRevD.104.022004} is $\omgw \leq 3.4\cdot10^{-9}$ at 25~Hz for a power law background with a spectral index of $2/3$ (consistent with expectations for CB mergers). This limit is very likely to improve with the next observing runs of LVK and, in the more distant future, with 3G detectors (Einstein Telescope, Cosmic Explorer) \citep{maggiore_science_2020,2021arXiv210909882E}.

Models of SGWB from stellar-mass CBs have been studied in recent years with different methods, in particular using extrapolations of the local observed merger rate and BH mass distribution \citep[e.g.][]{2016PhRvL.116m1102A,PhysRevD.98.063501,2021MNRAS.506.3977M,2021arXiv211105847L,babak2023stochastic}, analytical descriptions of BH formation and evolution via different channels and including the effects of the metallicity of progenitor stars
\citep[e.g.][]{dvorkin_metallicity-constrained_2016,Nakazato_2016,PhysRevD.100.063004,Mangiagli_2019} as well as detailed population synthesis models \citep[e.g.][]{perigois_startrack_2021,perigois_gravitational_2022}.

In this article we explore the SGWB from Binary Black Holes (BBHs) and Binary Neutron Stars (BNSs), with a focus on the frequency ranges accessible to LIGO/Virgo and LISA and using both analytical and population synthesis models. We note that the recent work of \citet{babak2023stochastic} has also addressed the SGWB from stellar-mass binaries in the LISA band. Our results, while using different astrophysical models, are in agreement with their conclusions, as we show below. Our goal here is to estimate the main sources of uncertainty stemming from astrophysical modelling and estimate the prospects of detection with LIGO/Virgo and LISA.

The structure of the article is as follows: Sec.~\ref{sec:simu} details all the ingredients we use to model the BBH and BNS populations, including mass, redshift, time delay and metallicity distributions. Sec.~\ref{sec:COSMIC} establishes our population synthesis models based on the COSMIC code \citep{COSMIC}. Sec.~\ref{sec:calcul} presents the calculation of the SGWB in  the LIGO/Virgo and LISA frequency bands. Sec.~\ref{sec:Nspace} discusses the possibility for LISA to detect individually some of these sources. Finally, we discuss our results and implications for future work in Sec.~\ref{sec:results}.

Throughout this article we use the following cosmological parameters: $h_0 = H_0/H_{100} = 0.68$ where $H_{100} = 100$ km s$^{-1}$ Mpc$^{-1}$, $\Omega_\Lambda = 0.69$, $\Omega_\mathrm{m} = 0.31$ \citep{2020A&A...641A...6P}.

\section{Modeling the BBH and BNS populations}\label{sec:simu}

This section describes all the components of our analytic modeling of the BBH and BNS populations. We leave the description of the population synthesis models to section \ref{sec:COSMIC}.

 Each CB is represented by a set of 6 parameters: the masses $M_1$ and $M_2$ of the binary components (described in sec.~\ref{subsec:mass}), the merger redshift of the CB (detailed in sec.~\ref{subsec:z}), its sky position represented by the right ascension $\alpha$ and the declination $\delta$, and finally the orbital inclination angle $\theta$. These three angles are drawn uniformly from the interval $[0, 2\pi]$ for $\alpha$, and from $[-1, 1]$ for $\cos{\delta}$ and $\cos{\theta}$. We regroup as $\lambda$ five parameters: $\lambda = \{M_1, M_2, \alpha, \delta,  \theta\}$.\\
The spins $\chi_1$ and $\chi_2$ of the binary components are assumed to be zero as their effect is subdominant for the total energy in GWs emitted in the inspiraling phase \citep{zhou_subtracting_2022}.

We assume that all CBs have circularized their orbits prior to entering the final phase of the inspiral. We note that while this assumption holds in the LIGO/Virgo band, this is not necessarily the case for LISA. Indeed, in some dynamical formation channels CBs enter the LISA band while still retaining some eccentricity, see for example \citet{Breivik:2016ddj}.

\subsection{Mass distributions}\label{subsec:mass}

\subsubsection{Black holes binaries}

We introduce the following probability distributions to describe the masses of the binary components: $P(M_1)$ for the mass of the primary and $P(q)$ for the mass ratio $q=M_1/M_2$. 
In order to model these distributions we use the LVK analysis of the third Gravitational-Wave Transient Catalog (GWTC-3) \citep{the_ligo_scientific_collaboration_population_2022}, based on the 69 confident BBH events that have a False Alarm Rate (FAR) below $1\, \mathrm{yr}^{-1}$.

We have chosen to use mainly their \textsc{Powerlaw+peak} (PL+P) model, as it provides the highest Bayes factor among the models considered in the catalog. We also discuss the impact of choosing other mass distributions, such as their \textsc{Powerlaw} (PL) and \textsc{Broken Power law} (BPL) models. These models are described in the appendix of \citet{the_ligo_scientific_collaboration_population_2022}.

In the PL+P model \citep{Talbot:2018cva} the distribution of the primary mass $M_1$ has two components: a power-law between $m_\mathrm{min}$ and $m_\mathrm{max}$, and a Gaussian peak with mean $\mu_\mathrm{m}$ and width $\sigma_\mathrm{m}$. The parameter $\lambda_\mathrm{peak}$ determines the fraction of the BBH systems which are contained in the Gaussian peak. The full list of parameters and their best-fit values are given in \citet{the_ligo_scientific_collaboration_population_2022}.

The $m_\mathrm{max} \sim 80\,\msol$ cut-off is motivated by the pair instability supernova (PISN) phenomenon. 
PISN are thought to occur in very massive ($m_\mathrm{ZAMS} \gtrsim 130 \msol$) and low metallicity ($Z \lesssim 0.1 Z_\odot$) stars \citep{2001ApJ...550..372F,2002ApJ...565..385U}, in which electron-positron pair creation leads to a thermonuclear runaway that completely disrupts the star, leaving no remnant \citep{2001ApJ...550..372F,2021ApJ...912L..31W}. Previous calculations have suggested that this effect may create a mass gap in the BH mass distribution in the range of $50-130$ $M_\odot$ \citep{2021ApJ...912L..31W}. The existence of this mass gap has been challenged by recent LIGO/Virgo observations that show BBH merging within the gap \citep{2020PhRvL.125j1102A,2020ApJ...900L..13A}.

The Gaussian peak at $\mu_\mathrm{m}\sim 35\,M_\odot$ could be explained by a similar phenomenon, but in less massive stars. In this case, the pair instability is not strong enough to completely disrupt the star, but can still lead to a transient regime where the star ejects matter to regain stability in a pulsating manner until its core eventually collapses \citep{1964ApJS....9..201F,2002ApJ...565..385U}.This pulsating pair-instability supernova (PPISN) leaves a remnant BH, which is however less massive than it would have been in the absence of the pulsating mechanism.

Finally, the secondary mass $M_2$ is calculated from the mass ratio $q$, which has a probability density that is a power law with a smoothed cut at $m_\mathrm{min}$ \citep{the_ligo_scientific_collaboration_population_2022}.

\subsubsection{Neutron stars binaries}

The mass distribution of BNSs is obtained from Galactic observations and assumed to be valid at all redshifts. As per the study conducted by  \citet{farrow_mass_2019}, this distribution can be accurately represented by a Gaussian with a mean of $1.33 \,M_\odot$ and a standard deviation of $0.09\, M_\odot$. We note that both neutron stars in the binary system are assumed to follow this mass distribution. The mass distribution of BNSs is not a critical parameter in this study as it is highly concentrated around a single value close to the Tolman–Oppenheimer–Volkoff limit, although this value is somewhat uncertain as it depends on the equation of state \citep{PhysRevLett.121.161101,doi:10.1146/annurev-astro-081915-023322,LATTIMER2016127}. 

It is important to note that the mass distribution of BNSs detected through GWs may differ from those observed in the Galaxy. This effect could indeed explain the observation of the BNS merger GW190425,  \citep{2020ApJ...892L...3A,2020MNRAS.496L..64R} whose total mass is much larger than that of Galactic binaries. Additionally, the recent observation of  GW190814 seems to have provided evidence for a BHNS merger with a NS much heavier than typical Galactic NSs \citep{the_ligo_scientific_collaboration_gw190814_2020}.

\subsection{Redshift distributions}\label{subsec:z}

To compute the merger rate of CBs, we first use the star formation rate (SFR) $\psi$, as given in \citet{vangioni_impact_2015} using a fit to observed data: 
\be
\label{eq:sfr}
\psi(z) = \nu \frac{a\,\exp{[b(z-z_m)]}}{a-b+b\,\exp{[a(z-z_m)]}}\,,
\ee
with $\nu = 0.178\, \msol\mathrm{Mpc}^{-3}\mathrm{yr^{-1}}$, $a = 2.37$, $b = 1.80$, $z_m = 2$.

We assume that the merger rate follows the SFR with some delay ($t_d$) which represents the time between the formation of the stellar progenitors
and the merger of the CB. We also consider the metallicity $Z$ which plays a key role in the evolution of the binary and thus affects the merger rate, as detailed in sec.~\ref{subsec:metal}. Thus, we can write the merger rate as: 

\be
\label{eq:Rmerg}
R_\mathrm{merg} (t) =  \int^{Z_\mathrm{max}}_{0} \int^{t_{d,\mathrm{max}}}_{t_{d,\mathrm{min}}}\alpha(Z)\, \psi(t-t_d)\,P(t_d|Z)\,P(Z|t-t_d)\,\mathrm{d}t_d\,\mathrm{d}Z\,,
\ee
where $\alpha$ is the efficiency (in $M_\odot^{\,-1}$) of forming CB that merge within the age of the Universe. Although the general redshift dependence of this efficiency is not known, we can use the local merger rate of BBHs and BNSs measured by LIGO/Virgo \citep{the_ligo_scientific_collaboration_population_2022} to constrain its value, as described below in section \ref{subsec:Rmerg0}.
Lastly, we take $Z_\mathrm{max} = Z_\odot = 0.014$, $t_{d,\mathrm{max}} = t_\mathrm{Hubble} = 13.8$ Gyrs and $t_{d,\mathrm{min}} = 10$ Myrs.

Note that, the time variable $t$ in eq.~(\ref{eq:Rmerg}) is connected to the redshift $z$ in eq.~(\ref{eq:sfr}) by :
\be
\frac{dt}{dz} = \frac{H_0^{-1}}{(1+z)\sqrt{\Omega_\mathrm{m} (1+z)^3 + \Omega_\Lambda}}\,.
\label{eq:tz}
\ee

In the following sections, we discuss each component of equation (\ref{eq:Rmerg}). 

\subsection{Time delay distributions}\label{subsec:td}

The time delay $t_d$ between the formation of the stellar progenitors and the merger of the CB can depend on their formation channel. It is usually assumed that the distribution of time delays is represented by a power-law probability function \citep[e.g.][see the discussion in Section \ref{sec:simu}]{2018MNRAS.474.2937C}. In equation (\ref{eq:Rmerg}), we consider the possible dependence of $t_d$ on the metallicity of the progenitor stars, which plays an important role in the formation of CB through its influence on the strength of the stellar winds. 
These processes are taken into account in population synthesis codes, which we discuss below. 

In the following, we consider analytic models in which $t_d$ does not depend on metallicity. Specifically, in the \textsc{baseline} model all time delays are taken to be zero, so that the CB merger rate follows the SFR, while in the \textsc{baseline\_delays} model the time delays are described by the probability distribution:
\be
\label{eq:Ptd}
P(t_d) \propto t_d^{-1}\,.
\ee
This functional form closely follows the results of population synthesis codes, as seen in Fig.~\ref{fig:TdCOSMIC} below.
This power law is restricted between $10$ Myrs and the Hubble time $t_\mathrm{Hubble}$.

Note that in both the \textsc{baseline} and the \textsc{baseline\_delays} model, the efficiency $\alpha$ in eq.~(\ref{eq:Rmerg}) is chosen to be constant.

\subsection{Metallicity distributions}\label{subsec:metal}

Metallicity affects the formation and evolution of BBHs in several ways. Firstly, higher metallicity can result in more efficient stellar winds, which can decrease the final masses of the progenitor stars and affect the properties of the BBHs formed~\citep{Chruslinska:2018hrb,Neijssel_2019}. Secondly, metallicity can impact the efficiency of mass transfer phases during binary star evolution (like the common envelope phase), which can result in less efficient mergers and a lower probability of forming BBHs~\citep{Neijssel_2019}. 
Finally, metallicity influences the properties of supernovae that produce BHs through fallback \citep{wong_fallback_2014}. In low-metallicity environments, the explosion mechanism is more likely to be asymmetric, leading to a larger fraction of black holes that receive a natal kick and thus breaking the binary. Additionally, the amount of mass ejected during a supernova is lower in low-metallicity environments, leading to the formation of more massive black holes \citep{spera_mass_2015,Belczynski2010}.

Metallicity evolves with cosmic time as stars form and then enrich their surroundings with metals both during and at the end of their lifetimes. However, the metallicity distribution in the Universe is not homogeneous, with a large dispersion at any given redshift between various galaxies and even within individual galaxies. In our analytic models, at a given redshift we nonetheless use averaged quantities over the entire galactic population. 

We take the metallicity evolution model from \citet{belczynski_first_2016}: 

\be
\label{eq:metalmean}
\overline{Z}(z) = \frac{y(1-R)}{\rho_b} \int_z^{z_\mathrm{max}} \frac{10^{0.5}\,\psi(z')}{H_0 (1+z')\sqrt{\Omega_\mathrm{m}(1+z')^3+\Omega_\Lambda}} \mathrm{d}z'\,,
\ee
where $R = 0.27$ is the return fraction (the mass fraction of each generation of stars that is ejected back into the interstellar medium),  $y=0.019$ is the net metal yield (the mass of new metals created and ejected into the interstellar medium by each generation of stars per unit mass locked in stars) and  $\rho_b = 2.77\cdot10^{11}\,\Omega_\mathrm{b} h_0\,\,\msol\,\mathrm{Mpc}^{-3}$  is the baryon density, with $\Omega_\mathrm{b} = 0.045$.

We also include a dispersion in metallicity at any given redshift. Assuming for simplicity this dispersion is log-normal  \citep{Santoliquido2021TheCM}, we obtain the following probability distribution of metallicity $Z$ at redshift $z$:
\be
\label{eq:metaldistrib}
P(Z|z) = \frac{1}{\sqrt{2\pi\sigma^2}} \,\,\exp{\left(-\frac{(\log{(Z/Z_\odot)} - \log{(\overline{Z}(z)/Z_\odot)})^2}{2\sigma^2}\right)}\,,
\ee
with $\sigma = 0.2$. The  probability $P(Z|t-t_d)$ that we use in eq.~(\ref{eq:Rmerg}) is then obtained using the redshift-time mapping of eq.~(\ref{eq:tz}).

If we make the assumption that the probability $P(t_d|Z)$ in eq.~(\ref{eq:Rmerg}) does not depend on the metallicity, for example if it is a power law as considered in Section \ref{subsec:td}, then we can factorize the two integrals in the merger rate equation to obtain:
\be
\label{eq:Rmerg2}
R_\mathrm{merg} (t) = \int^{t_{d,\mathrm{max}}}_{t_{d,\mathrm{min}}} \psi(t-t_d)\,P(t_d)\,dt_d \, \times \, f_Z(t-t_d)\,,
\ee
with the merging fraction $f_Z(z(t))$:
\be
\label{eq:fZ}
f_Z(z(t)) = \int^{Z_\mathrm{max}}_{0} \alpha(Z) \, P(Z|z)\,\mathrm{d}Z \,.
\ee

We assume that no BHs could be formed by a progenitor star with a metallicity above 10\% of the solar one. This cut is inspired by population synthesis results \citep[see e.g.][]{Santoliquido2021TheCM,Srinivasan+23}. As a result, under our assumptions, $\alpha$ is a step function centered on $Z_\mathrm{cut} = 0.1 \,Z_\odot$. Then $f_Z(z)$ is a step-like function with a cut-off around $z = 4.5$. The cut-off is sharpened with the decrease of $\sigma$ in eq.~(\ref{eq:metaldistrib}).

These metallicity effects are taken into account in our \textsc{metallicity cut} model. Specifically, we use eq.~(\ref{eq:Rmerg2}) to calculate the merger rate with the metallicity distribution given in eq.~(\ref{eq:metaldistrib}) and the time delay distribution from eq.~(\ref{eq:Ptd}). Note that for this model the time delay distribution is the same as for the \textsc{baseline\_delays} model, but the efficiency of forming CBs now depends on redshift.

\subsection{Local rates and astrophysical uncertainty}\label{subsec:Rmerg0}

In our analytical models, we do not know a priori the general redshift dependence of the efficiency $\alpha$ in eq.~(\ref{eq:Rmerg}). However, we can constrain its value by using the local merger rate obtained from the GWTC-3 catalog by LIGO-Virgo \citep{the_ligo_scientific_collaboration_population_2022}. That is, we take $\alpha$ constant in order that locally, the merger rate given in eq.~(\ref{eq:Rmerg}) matches the value measured by LIGO/Virgo.

For BBHs, we use the rate provided by the \textit{PP} model in GWTC-3, which is $R_\mathrm{BBH}(z=0.2)=28.3^{+13.9}_{-9.1}$ yr$^{-1}$Gpc$^{-3}$. This rate is given for $z=0.2$ since it corresponds to the redshift where the uncertainty is minimal. For BNSs, we use the rate provided by the \textit{PDB (pair)} model in GWTC-3, which is $R_\mathrm{BNS}(z=0)=44^{+96}_{-34}$ yr$^{-1}$Gpc$^{-3}$. The uncertainties quoted for these values correspond to 90\% confidence intervals.

Note that, in our analytical models, we have assumed that the relevant quantities, such as the SFR, metallicity, time delay, mass, and redshift distribution, have already been averaged over the mass distribution of the host galaxies. This implies that the host galaxy dependence has been accounted for by taking an average over all possible hosts. This simplifies the analysis by eliminating the need to integrate the merger rate over the host galaxy mass distribution.
However, we do not make this assumption for our BBH population synthesis model, as we detail in the next section.

\section{Population synthesis models}\label{sec:COSMIC}

We supplement the analytic models described above with a set of population synthesis models that rely on a more realistic description of stellar evolution.
We follow the prescription of \citet{Srinivasan+23} to produce an astrophysical population of BBHs that merge by the present day. We account for the SFR through cosmic time as a function of progenitor environment properties (metallicity and galaxy mass) and the efficiency of binary black hole formation from binary stars. The former is adopted from the default star formation rate of \citet{Srinivasan+23} that models the mass-metallicity relation based on the metallicity calibration of \citet{KK04} and extrapolated to high redshift based on the fit provided by \citet{Ma+2016}. The latter is obtained using a rapid binary population synthesis code, COSMIC (v3.4.0) \citep{COSMIC}, which simulates the binary evolution of stars that ultimately form merging binary black holes. These binary simulations are based on parametric models of single-star evolution and their binary interactions. We adopt the default binary evolution parameters shown in bold in Table 1 of \citet{Srinivasan+23} to generate a population of merging binary black holes, \textsc{BBH COSMIC default}. 

Uncertainties in modeling the SFR and binary evolution result in a large variance in the predicted merger rate. Therefore, we explore models with a pessimistic and optimistic merger rate in comparison to the \textsc{BBH COSMIC default}, termed \textsc{BBH COSMIC pessimistic} and \textsc{BBH COSMIC optimistic}, respectively. \textsc{BBH COSMIC pessimistic} differs from the default model in two aspects: a different critical mass ratio model of the onset of unstable mass transfer (qcflag = 4 as per COSMIC documentation \footnote{https://cosmic-popsynth.github.io/} and a different common-envelope efficiency $\alpha = 0.1$). Likewise, \textsc{BBH COSMIC optimistic} varies in the description of the common envelope phase  (lambdaf = 0 and $\alpha = 0.2$ as per COSMIC documentation). The present-day merger rates of our pessimistic, default, and optimistic models are 15, 120, 590 $\mathrm{Gpc^{-3}\,yr^{-1}}$ respectively. \\

For the BNSs, we take the \textsc{default} COSMIC parameters of \citet{Srinivasan+23}. To compute the merger rate, we use equation (\ref{eq:Rmerg}) with SFR and metallicity distribution $P(Z|z)$ provided respectively in eq.~(\ref{eq:sfr}) and eq.~(\ref{eq:metaldistrib}). We use 15 log-uniform  metallicity bins between $4.4\cdot10^{-4}$ and $2.8\cdot10^{-1}$. The mass efficiency $\alpha(Z)$ and the time delay distribution $P(t_d|Z)$ are calculated from our COSMIC results.\\

All of our models are summarized in Table \ref{tab:models}. Analytical models (\textsc{baseline}, \textsc{baseline\_delays}, \textsc{metallicity cut}) are easy to calibrate and to use because of their small number of parameters. Each new analytical model adds a physical process (time delay, metallicity) that refines the model and makes it more realistic. Nonetheless, these analytical models are not detailed enough and cannot capture some aspects of BBHs and BNSs formation. That is why we choose to compare them with more realistic models based on the population synthesis code COSMIC, as outlined above. This code describes better the physical processes at play during stellar evolution, taking into account several physical effects (common envelope phase, kick velocity etc. see \citet{COSMIC}) and it is in this sense more realistic. On the other hand, population synthesis codes have numerous parameters that are difficult to constrain, see \citet{2022LRR....25....1M} for a recent review. Looking at observables produced by a set of COSMIC parameters (for example BBHs merger rate and stochastic backgrounds) and comparing them to observations could help us to calibrate COSMIC parameters. In practice, there are currently not enough observational constraints to overcome the degenerates between various COSMIC parameters. In this work,  we are not attempting to calibrate COSMIC parameters, rather we use specific sets of parameters that represent well the range of variability of COSMIC.\\

\begin{figure}
\centering
\includegraphics[scale=0.56]{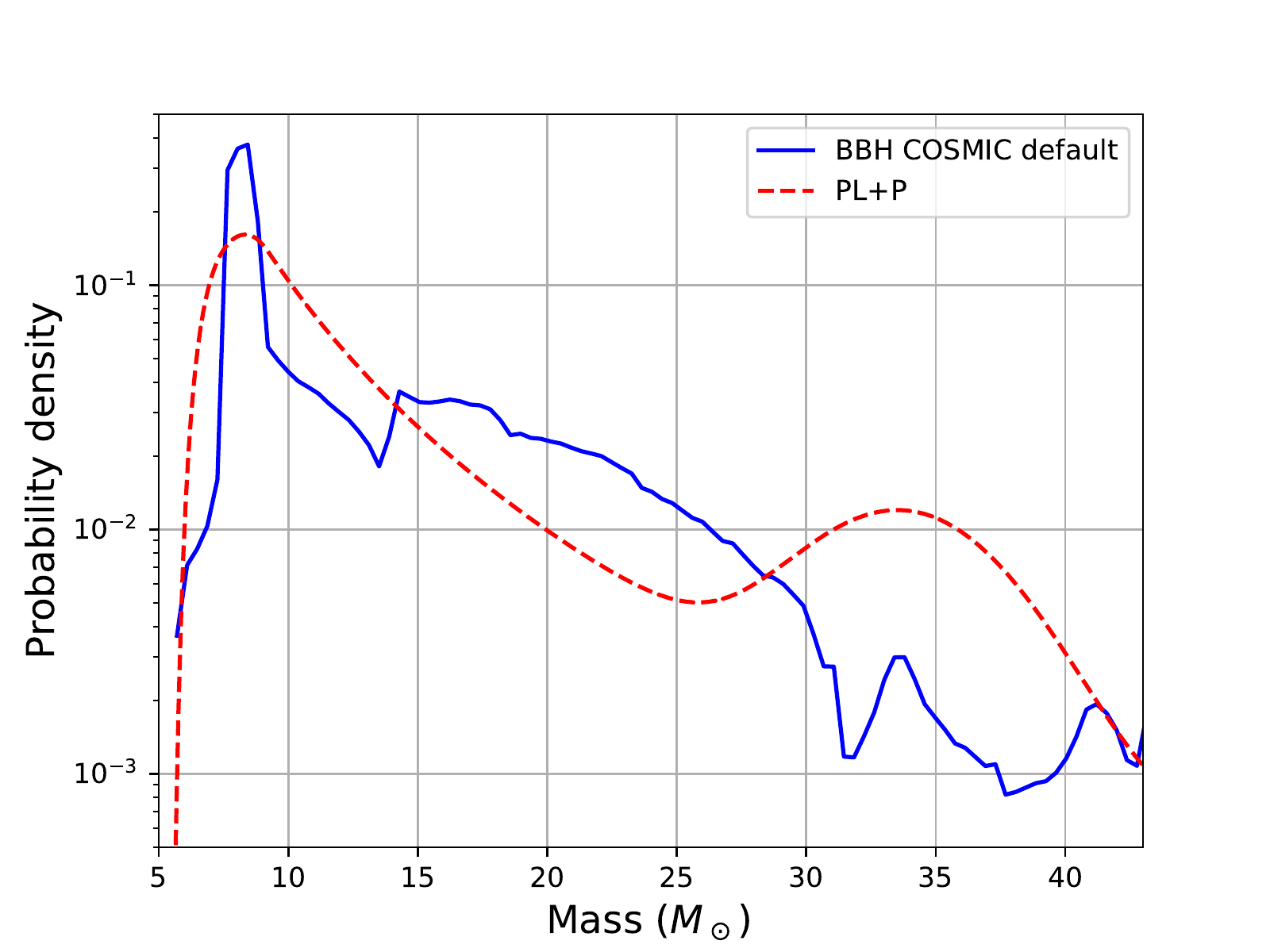}
\caption{The distribution of the primary mass $M_1$ for the \textsc{Powerlaw+peak} model from the GWTC-3 catalog and our COSMIC simulations with default parameters. The peak at around $M_1 \sim 10M_{\odot}$ as well as the power-law-like behavior are similar in both models. Note however the differences at high masses, at $M_1 \gtrsim 30M_{\odot}$.}
\label{fig:massCOSMIC}
\end{figure}

\begin{figure}
\centering
\includegraphics[scale=0.56]{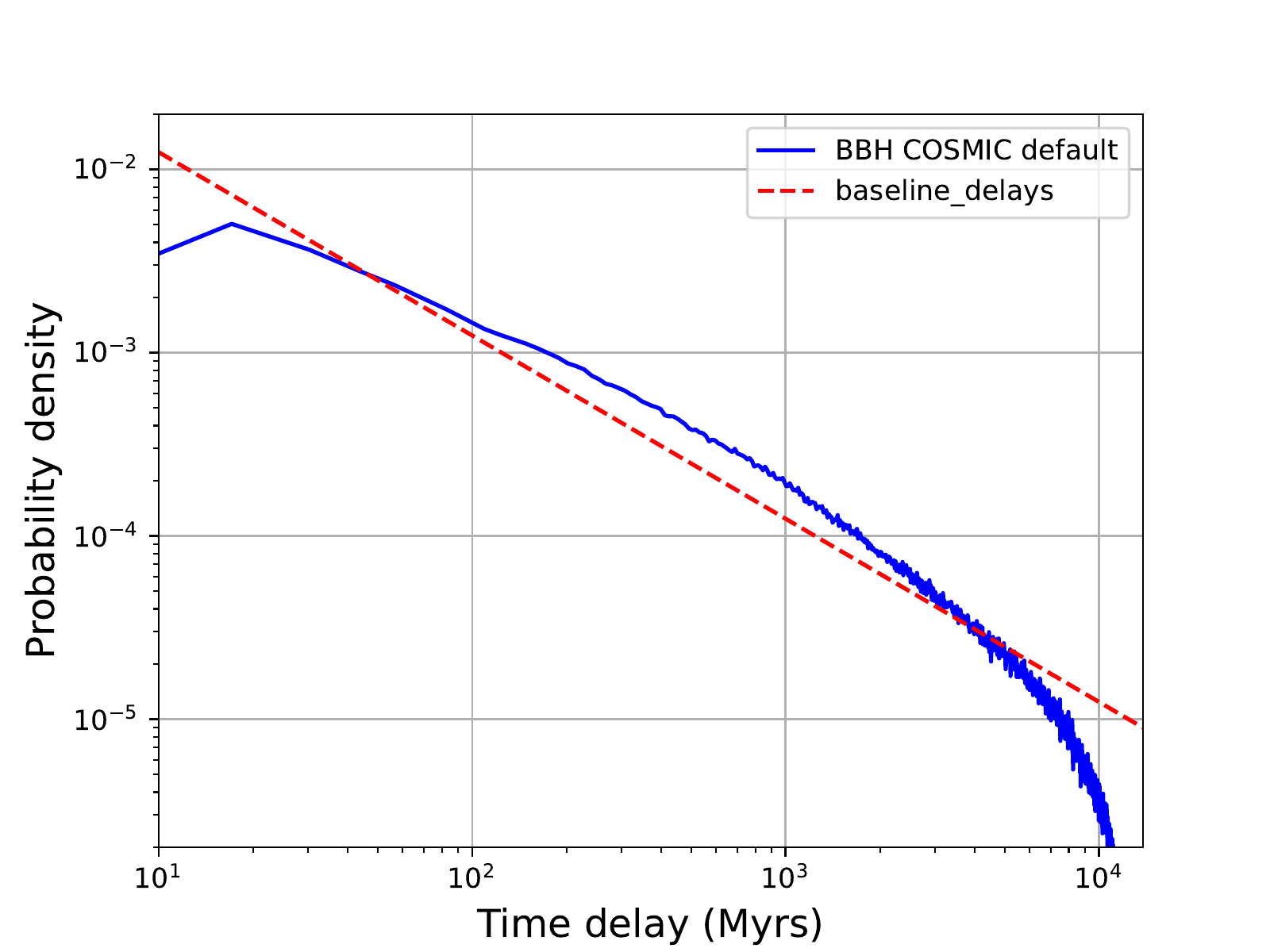}
\caption{Time delay probability density integrated over all metallicities for our analytical \textsc{baseline\_delays} model and our COSMIC simulation with the default parameters. The power-law time delay model adopted in our analytic prescriptions provides a good description of the detailed COSMIC model.}
\label{fig:TdCOSMIC}
\end{figure}

In Figure \ref{fig:massCOSMIC} we show the BBH primary mass distribution of the PL+P model and the one resulting from our default COSMIC model. The COSMIC distribution has a more pronounced peak at low masses around $8\,M_\odot$ and presents an over density centered around $22\,M_\odot$ but does not have any peak around $34\,M_\odot$ contrary to the PL+P model.

Figure \ref{fig:TdCOSMIC} compares the time delay probability density integrated over all metallicities in our analytical \textsc{baseline\_delays} model and our default COSMIC model. The distribution from COSMIC is very close to a power law but with a turn-over at very short time delays.\\

\begin{figure}
\centering
\includegraphics[scale=0.56]{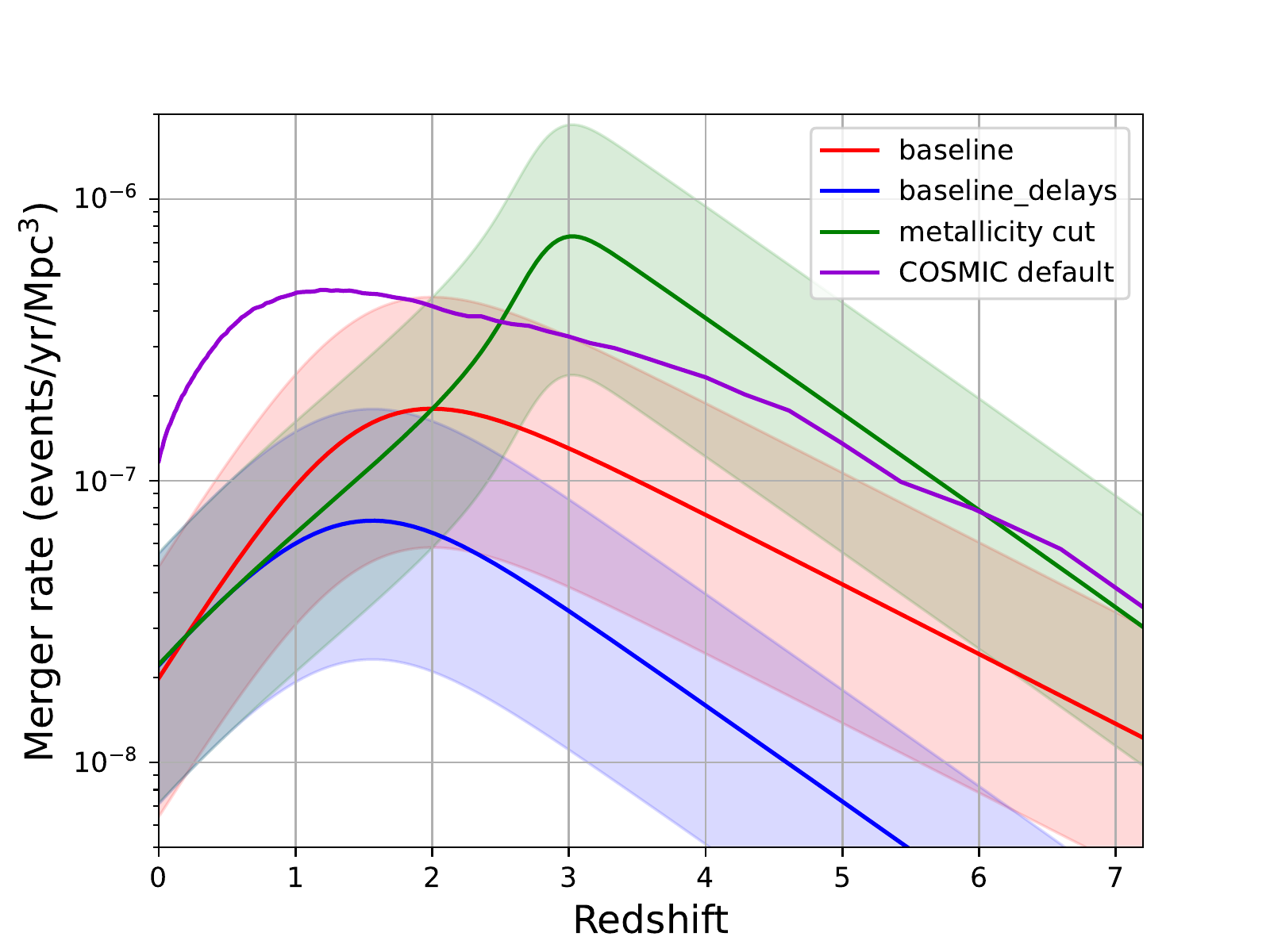}
\caption{The merger rate of BBHs calculated with our main models. The uncertainties in this figure come from the 90\% confidence interval on the merger rate at $z=0.2$ from the GWTC-3 catalog. We assume the uncertainty to be the same at every redshift.}
\label{fig:Rmerg}
\end{figure}

We compare the merger rate of BBHs in our models in Figure \ref{fig:Rmerg}. As expected, since there is no time delay between formation and merger, the shape of the \textsc{baseline} merger rate is close to the SFR [see eq.~(\ref{eq:sfr})] and peaks at around $R_\mathrm{merg}(z=2) \sim 1.5\cdot 10^7$ events/yr/Mpc$^3$.

We also observe that for $z > 0.2$ the \textsc{baseline\_delays} merger rate is always below the \textsc{baseline} model. The reason is that the inclusion of time delays shifts the entire \textsc{baseline} merger rate curve to lower redshifts. Then this shifted curve needs to be renormalized to be in accordance with the local LIGO/Virgo observations at $z=0.2$, as described in section \ref{subsec:Rmerg0}. This procedure then results in a merger rate that is below the \textsc{baseline} model at all redshifts above $z=0.2$.

The merger rate in the \textsc{metallicity cut} model is slightly lower than the \textsc{baseline} one at small redshifts ($z<2$), but significantly higher at larger redshifts ($z>2$). This is due to the fact that the effect of metallicity is to reduce the fraction of massive stars that can collapse into BHs. At lower redshifts, stars tend to have higher metallicity, resulting in fewer stars collapsing into black holes. Conversely, at higher redshifts, stars tend to have lower metallicity, leading to a larger fraction of stars collapsing into black holes. The asymmetry of this trend around the \textsc{baseline} model results predominantly from the renormalization of the merger rate at low redshifts ($z=0.2$).

Finally, the merger rate from the \textsc{COSMIC default} model is significantly higher than our other models and peaks at lower redshift around $z=1$. This model predicts a merger rate at $z = 0.2$ which is outside the 90\% confidence interval computed in the GWTC-3 catalog \citep{the_ligo_scientific_collaboration_population_2022}. This implies that the default set of parameters from \citet{Srinivasan+23} does not describe the observed merger rate of BBHs. Note that this difference in merger rate can be also due to our model for star formation and metallicity evolution.

In this study, we choose to explore the range of uncertainties due to COSMIC parameters rather than trying to find the best parameter set. For this reason, we include also two other sets (\textsc{pessimistic} and \textsc{optimistic} in Table \ref{tab:models}) from the same study that represent well the range of variability of COSMIC models. The former set has a merger rate at $z=0.2$ that is in accordance with GWTC-3 while the latter provides an upper bound on the merger rate that COSMIC can predict.

\begin{table}
\begin{tabular}{|c|ccc|}
\hline
Models name                       & \multicolumn{1}{c|}{BBHs mass dist.} & \multicolumn{1}{c|}{Time delay dist.} & Metallicity \\ \hline
\textsc{baseline}         & \multicolumn{1}{c|}{PL+P}           & \multicolumn{1}{c|}{$\varnothing$}                 &      $\varnothing$       \\ \hline
\textsc{baseline\_delays} & \multicolumn{1}{c|}{PL+P}           & \multicolumn{1}{c|}{Power-law, eq.~(\ref{eq:Ptd})}               &      $\varnothing$       \\ \hline
\textsc{metallicity cut}  & \multicolumn{1}{c|}{PL+P}           & \multicolumn{1}{c|}{Power-law, eq.~(\ref{eq:Ptd})}              &     eqs.~(\ref{eq:Rmerg2}-\ref{eq:fZ})         \\ \hline \hline
COSMIC                     & \multicolumn{3}{p{\dimexpr0.75\linewidth-2\tabcolsep}|}{3 models: \textsc{default}, \textsc{pessimistic}, \textsc{optimistic}. They have different sets of parameters (described in sec.~\ref{sec:COSMIC}) that change the stellar evolution.} \\ \hline
\end{tabular}
\caption{Summary of our 3 analytical models and 3 population synthesis models. The PL+P BBHs mass distribution is taken from \citet{the_ligo_scientific_collaboration_population_2022}. The $\varnothing$ symbol means that the model does not take into account the related parameter.}
\label{tab:models}
\end{table}

\section{Calculation of the SGWB from BBH and BNS}\label{sec:calcul}

The total dimensionless energy density of gravitational waves $\omgw$, per logarithmic unit of frequency and in unit of the critical density of the Universe $\rho_c$, is expressed as:

\be
\omgw = \frac{1}{\rho_c} \frac{\mathrm{d}\rho_\mathrm{GW}}{\mathrm{d}\ln{f}}\,,
\label{eq:defOmGW}
\ee
with $\mathrm{d}\rho_\mathrm{GW}$ the energy density of the GWs in the frequency interval $[f, f +df]$.

We can write the background from BBHs or  BNSs mergers as: 
\be
\label{eq:OmGW}
\omgw(f) = \frac{f}{\rho_c c^2 H_0}\, \int^{z_\mathrm{max}}_0 \int_{\lambda} \,\,\frac{R_\mathrm{merg}(z,\lambda)\, \frac{\mathrm{d}E_\mathrm{GW}(f_s)}{\mathrm{d}f_s} \, P(\lambda) }{ (1+z)\, \sqrt{\Omega_M (1+z)^3 + \Omega_\Lambda}}\,\mathrm{d}\lambda \,\mathrm{d}z\,,
\ee
with $f$ the observed frequency, $f = f_\mathrm{s}/(1+z)$ the frequency emitted at the source, $R_\mathrm{merg}$ the CB merger rate, $\mathrm{d}E_\mathrm{GW}(f_s)/\mathrm{d}f_s$ the energy spectrum emitted by each CB and $P(\lambda)$ the probability distribution of the parameters of the CB, summarized as $\lambda$.

We used the phenomenological expression given in \citet{perigois_startrack_2021} for the energy spectrum emitted by each CB. The coefficients of this expression are obtained by matching post-newtonian and numerical relativity waveforms \citep[see the appendix and][]{ajith_inspiral-merger-ringdown_2011}.

\subsection{LIGO/Virgo and LISA sensitivity to a SGWB}\label{subsec:sensi}

We use the LIGO/Virgo sensitivity curves to a SGWB given by the LIGO/Virgo/Kagra collaboration \citep{the_ligo_scientific_collaboration_population_2022}. For LISA, we use an analytic approximation for the sensitivity curve for a point like source given in \citet{robson_construction_2019} to calculate the power-law integrated sensitivity to a SGWB.

We define the latter to be the limit at which the signal-to-noise ratio (SNR) for detecting a signal is equal to 5, assuming an observation time of $T_\mathrm{obs} = 4$ years. A reasonable frequency range to calculate the sensitivity of LISA is between  $10^{-1}-10^{2}$ mHz. To compute the power-law integrated sensitivity curve, we use the method proposed by \citet{thrane_sensitivity_2013}.

\section{Stellar-mass CB mergers detectable by LISA}\label{sec:Nspace}

Even though the mergers of stellar mass black holes do not emit their maximum intensity in the LISA band, it could be possible to detect these systems individually with LISA.
In fact, LISA is not well suited to detect stellar-mass BBH and BNS because its sensibility is in the mHz regime. Nonetheless, by accumulating the signal over multiple periods, the SNR can be increased and pass beyond the detection threshold.
We estimate the expected number of individual detections $N_\mathrm{LISA}$ with LISA for a mission duration of $T_\mathrm{obs} = 4$ years using the method described in \citet{gerosa_multiband_2019}: 

\be
N_\mathrm{LISA} = \int_z \int_{\lambda}  P(\lambda)\,R_\mathrm{merg}(z)\,\frac{\mathrm{d}V_c}{\mathrm{d}z}\,\frac{1}{1+z}\,\Delta(\lambda,z)\,\mathrm{d}z \,\mathrm{d}\lambda\,.
\label{eq:Nspace}
\ee

The quantity $\Delta(\lambda,z)$ provides an estimate of the time window in which a merging CB is visible by LISA with SNR above a threshold value SNR$_{thr}$. We chose SNR$_{thr}=8$ since the source parameters were shown to be well constrained in this case \citep{2021PhRvD.104d4065B}, although we note that lower values can be considered for sources observable with ground-base detectors \citep{2018PhRvL.121y1102W}.

We need to remove these sources from the background, since they are individually detected. Removing these sources will lower the background, thus some new sources could become individually detectable since the background contributes to the overall noise budget. As a result, $\Delta$ in eq. (\ref{eq:Nspace}) is likely to increase with the decrease of the background. Therefore, we need to recompute $N_\mathrm{LISA}$ and eventually repeat this process until we reach convergence. In practice, this iterative process is not necessary, indeed in our case $N_\mathrm{LISA}$ is of the order of magnitude of $10$ for the BBHs and zero for the BNSs (see the results section \ref{sec:results} below). Thus, detectable stellar-mass sources have a negligible contribution to the SGWB for LISA.

\section{Results}\label{sec:results}

\begin{figure}
\centering
\includegraphics[scale=0.56]{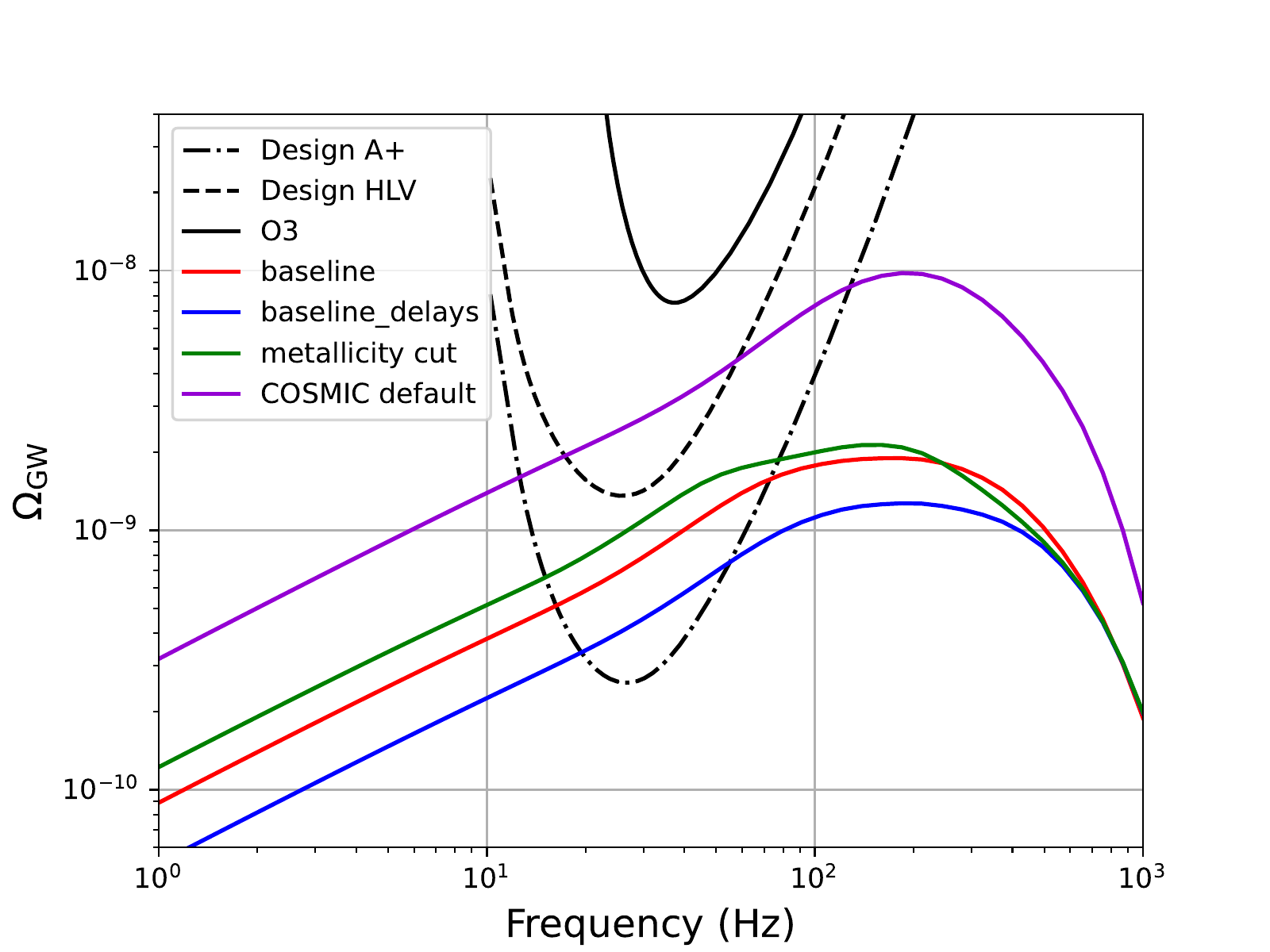}
\caption{In colored lines, the SGWB from BBHs calculated with our main models (see Table \ref{tab:models}). In black lines, the LIGO/Virgo SGWB sensibility. There is a factor 3 variability in the background predicted by our analytical models. The prediction obtained using COSMIC \textsc{default} significantly exceeds the analytical models, mainly due to its higher merger rate (see Fig.~\ref{fig:Rmerg}).}
\label{fig:overview}
\end{figure}

In this section, we present and analyze the results of the various models for the SGWB from BBH and BNS sources, in both the LISA and LIGO/Virgo frequency ranges. The main results are summarized in Figure \ref{fig:overview} and Table \ref{tab:overview}, which provides the SGWB values at 3~mHz and 25~Hz, which are the reference frequencies, respectively for LISA and LIGO/Virgo. All of our models are consistent with the upper limit from LVC O3 observations, but only the \textsc{cosmic} model predicts a SGWB strong enough to be confidently detected with design LIGO/Virgo sensitivity.

\begin{table}
\begin{tabular}{|c|c|c|c|c|}
\hline
Models                    & \textsc{baseline} & \textsc{baseline\_delays} & \textsc{Z cut} & \textsc{COSMIC} \\ \hline
$(25\,\mathrm{Hz})\,\, \omgw\cdot10^{10}$ & $6.83^{+3.35}_{-2.20}$ & $3.99^{+1.96}_{-1.28}$ & $9.42^{+4.63}_{-3.03}$ & 24.11 \\ \hline
$(3\,\mathrm{mHz})\,\,\omgw\cdot10^{12}$ & $1.89^{+0.93}_{-0.61}$ & $1.10^{+0.54}_{-0.35}$ & $2.61^{+1.28}_{-0.84}$ & 6.75 \\ \hline \hline
$N_\mathrm{LISA}$ & $6^{+3}_{-2}$ & $7^{+3}_{-2}$ & $7^{+3}_{-2}$ & 19 \\ \hline
\end{tabular}
\caption{Values of the SGWB in figure \ref{fig:overview} and the number of the individually detected BBH mergers with LISA with an SNR of at least 8 for a 4 years observation run. $N_\mathrm{LISA}\sim 10$ for all of our analytic models. The error bars on the background from analytical models come from the 90\% confidence interval on the merger rate at z = 0.2 (GWTC-3 catalog), on which these models are calibrated. Error bars can not be provided for population synthesis-based models, but in Figure \ref{fig:COSMIC} and in the text we discuss their range of uncertainty.}
\label{tab:overview}
\end{table}

Compared to the \textsc{baseline} model, the \textsc{baseline\_delays} takes into account the time delay of the CB mergers, leading to a decrease in the SGWB by a factor of $1.7$ across all frequencies. Indeed, as we showed in section \ref{sec:COSMIC}, the \textsc{baseline\_delays} merger rate is lower compared to the \textsc{baseline} at all redshifts, resulting in a lower background since the other parameters in eq.~(\ref{eq:OmGW}) remain unchanged for both of these models.

The \textsc{metallicity cut} model includes the effect of metallicity, resulting in an amplitude increase by a factor of $1.4$ and a shift of the peak towards lower frequencies. These effects can be explained
by the combination of two effects. First, as shown in Section \ref{sec:COSMIC}, the \textsc{metallicity cut} merger rate is higher than the \textsc{baseline} one for most of the redshift range, thus the corresponding background is higher. Second, as can be seen in Fig.~\ref{fig:Rmerg}, the merger rate for the \textsc{metallicity cut} model peaks close to  $z=3$. As a result, the GW frequencies of the background for this model are more redshifted.

Note that a second bump is observed in the \textsc{metallicity cut} merger rate around $50$~ Hz. As we show below, this feature is due to the peak in the PL+P mass distribution. 

The SGWB obtained for the \textsc{cosmic} model is higher than for the other models, and so upcoming observations in run O4 are expected to have sufficient sensitivity to either detect or place strong constraints on this model. It is important to mention that the peak around 200 Hz in the COSMIC model appears to be more prominent than in other cases. This is due to the mass distribution produced by the COSMIC simulation, which peaks at lower masses than the distribution presented in GWTC-3  \citep{the_ligo_scientific_collaboration_population_2022}, as shown in Figure \ref{fig:massCOSMIC}.

\begin{figure}
\centering
\includegraphics[scale=0.56]{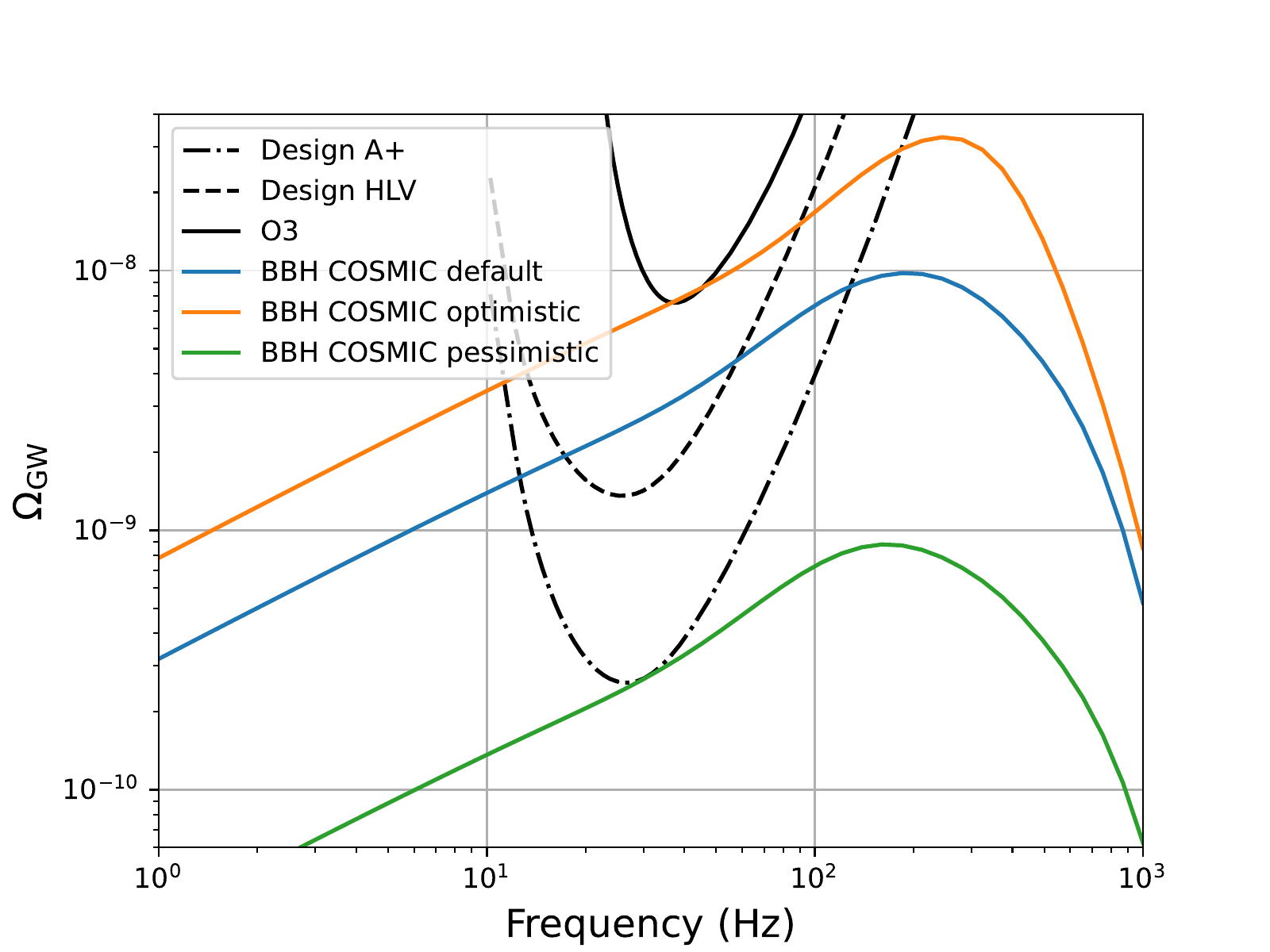}
\caption{The SGWB from BBHs, based on the results of our COSMIC simulations discussed in section \ref{sec:COSMIC}. There is a variation of one order of magnitude between our models. The COSMIC \textsc{optimistic} model seems to be already excluded by observation, since no background has been detected during O3.}
\label{fig:COSMIC}
\end{figure}

Figure \ref{fig:COSMIC} compares the results of the three different COSMIC runs described in section \ref{sec:COSMIC}. We choose these runs because they illustrate well the range of uncertainty of binary population synthesis models. It can be seen that changing the stellar evolution parameters of the COSMIC code leads to a variation of one order of magnitude in the resulting background, from the O3 limits to the design sensitivity. Therefore, the range predicted by the COSMIC models studied in this paper will be probed in the near future.
In fact, the \textsc{optimistic} model appears to be already excluded by observations, as no background has been detected during O3. Thus, this also excludes the merger rate predicted by this model. The upcoming O4 run is expected to constrain even more these parameters.

\begin{figure}
\centering
\includegraphics[scale=0.56]{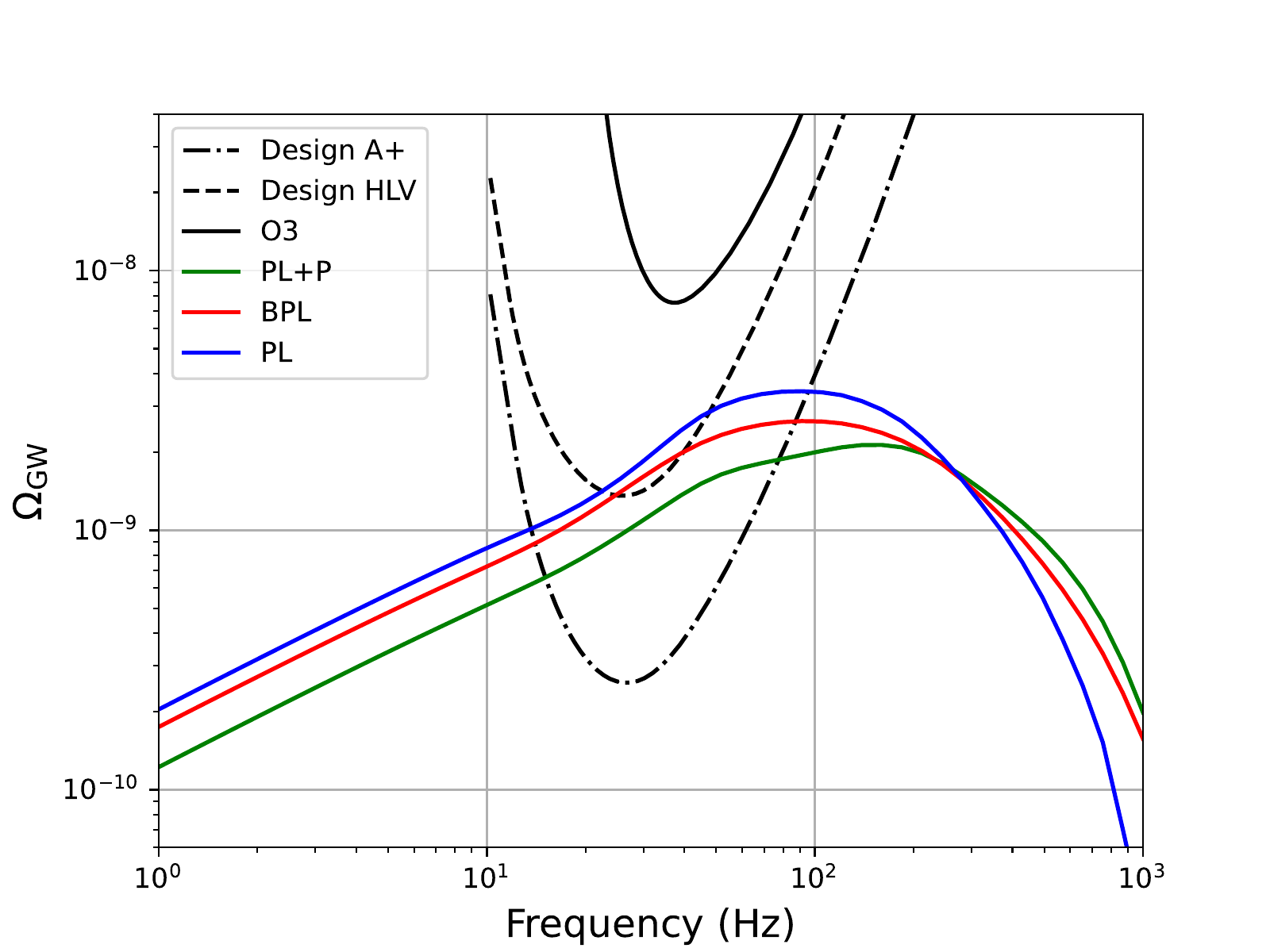}
\caption{The SGWB for BBHs mass distribution models described in \citet{the_ligo_scientific_collaboration_population_2022} and the \textsc{metallicity cut} model. Varying the mass distribution results in a factor $2$ difference of the SGWB.}
\label{fig:massDistrib}
\end{figure}

In Figure \ref{fig:massDistrib}, we set the model to be \textsc{metallicity cut}, and we examine the impact of the mass distribution model on the background. The PL+P distribution results in a broad peak with two distinguishable bumps around 50 and 200 Hz, which are a result of the two peaks in this mass distribution (around 5 and 35 $M_\odot$). On the other hand, the PL model gives only one peak around 100 Hz and a bit higher background overall. The BPL model is intermediate between the two, with a lower background but a wider peak. Upcoming O4 observations have the potential to constrain the parameters of the mass distribution models and thus the resulting background.

\begin{figure}
\centering
\includegraphics[scale=0.56]{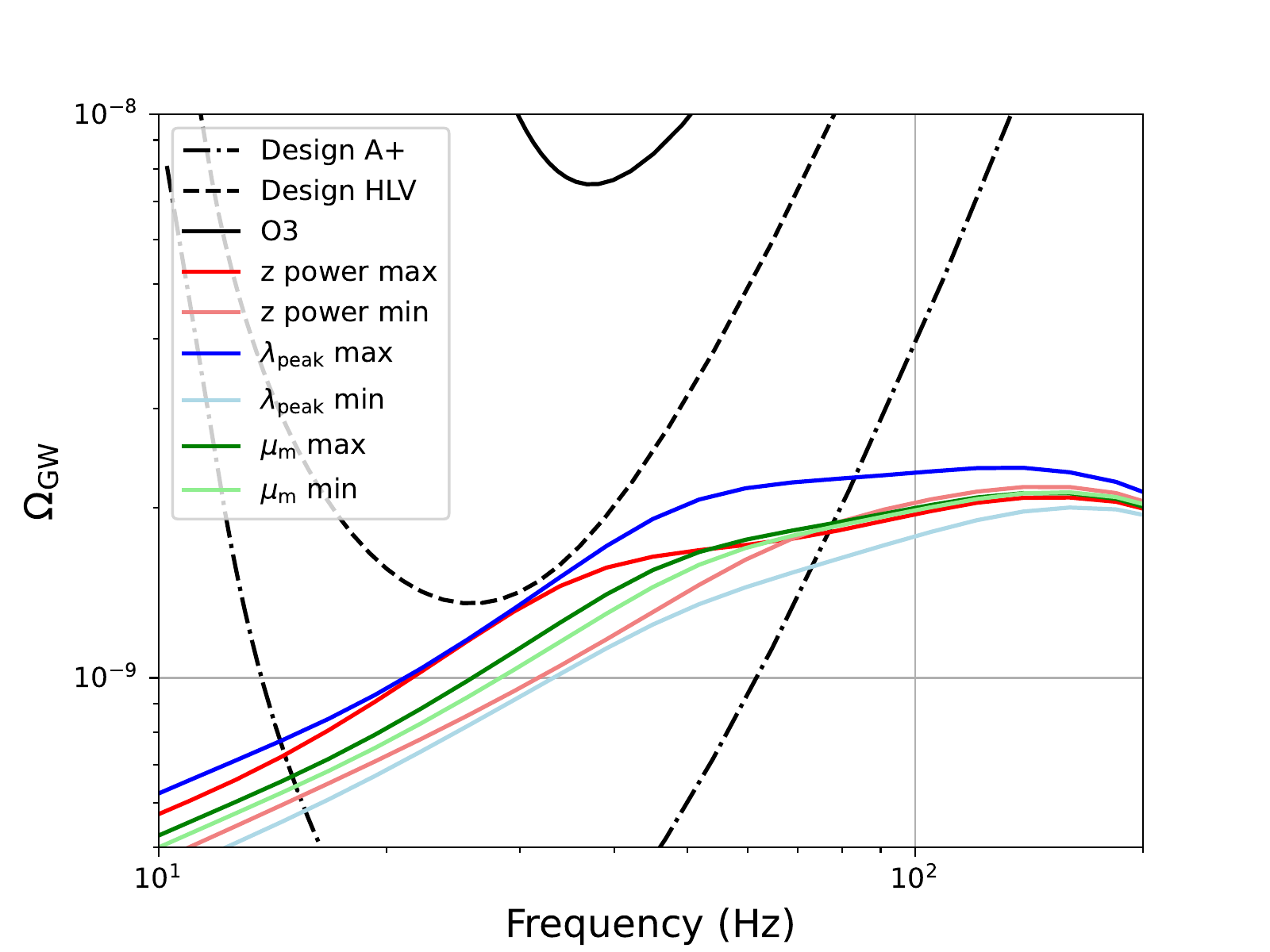}
\caption{The variation of the SGWB from BBHs due to uncertainties on the PL+P mass distribution. Each colored curve represents the predicted background by our \textsc{metallicity cut} model with each time a different value for one parameter of the PL+P distribution. The uncertainty on the amplitude of the gaussian peak of the PL+P model results in the most significant variation on the background.}
\label{fig:varParam}
\end{figure}

Fixing the mass model to be PL+P, we now study the impact of the main parameters of this mass distribution on the resulting background. To this end, we vary the amplitude ($\lambda_\mathrm{peak}$) and the position ($\mu_\mathrm{m}$) of the Gaussian peak in the mass distribution. These values are set to their maximum and minimum of the 90\% credibility interval, and the resulting backgrounds are compared in Figure \ref{fig:varParam}.

We also investigate the impact of varying $\mu_\mathrm{m}$ with redshift by taking $\mu_\mathrm{m} \propto (1+z)^{0.5}$ ("z power max" in Figure \ref{fig:varParam}) or $\mu_\mathrm{m} \propto (1+z)^{-0.5}$ ("z power min"). Indeed, $\mu_\mathrm{m}$ may vary with the redshift since, as we discuss in section \ref{subsec:mass}, the Gaussian peak in the mass distribution could be the result of PPISNs and the efficiency of this phenomenon depends on the metallicity (and so on the redshift).

The uncertainty in the amplitude of the peak results in the most significant variations of the SGWB. When set to its minimum, the bump around 50 Hz is almost imperceptible. The uncertainty in the absolute position of $\mu_\mathrm{m}$ results in only minor variations in the background. However, assuming a strong redshift dependence moves the bump from 35 Hz ("z power max") to 75 Hz ("z power min").

Note that the bump in the background around 200 Hz does not change. This confirms that the secondary bump around 50 Hz is indeed due to the peak around 35 $M_\odot$ in the PL+P mass distribution. Furthermore, this secondary bump is the closest part of the SGWB to the minimum sensitivity of LIGO/Virgo and thus could be one of the first features of the SGWB to be detected.

\begin{figure}
\centering
\includegraphics[scale=0.56]{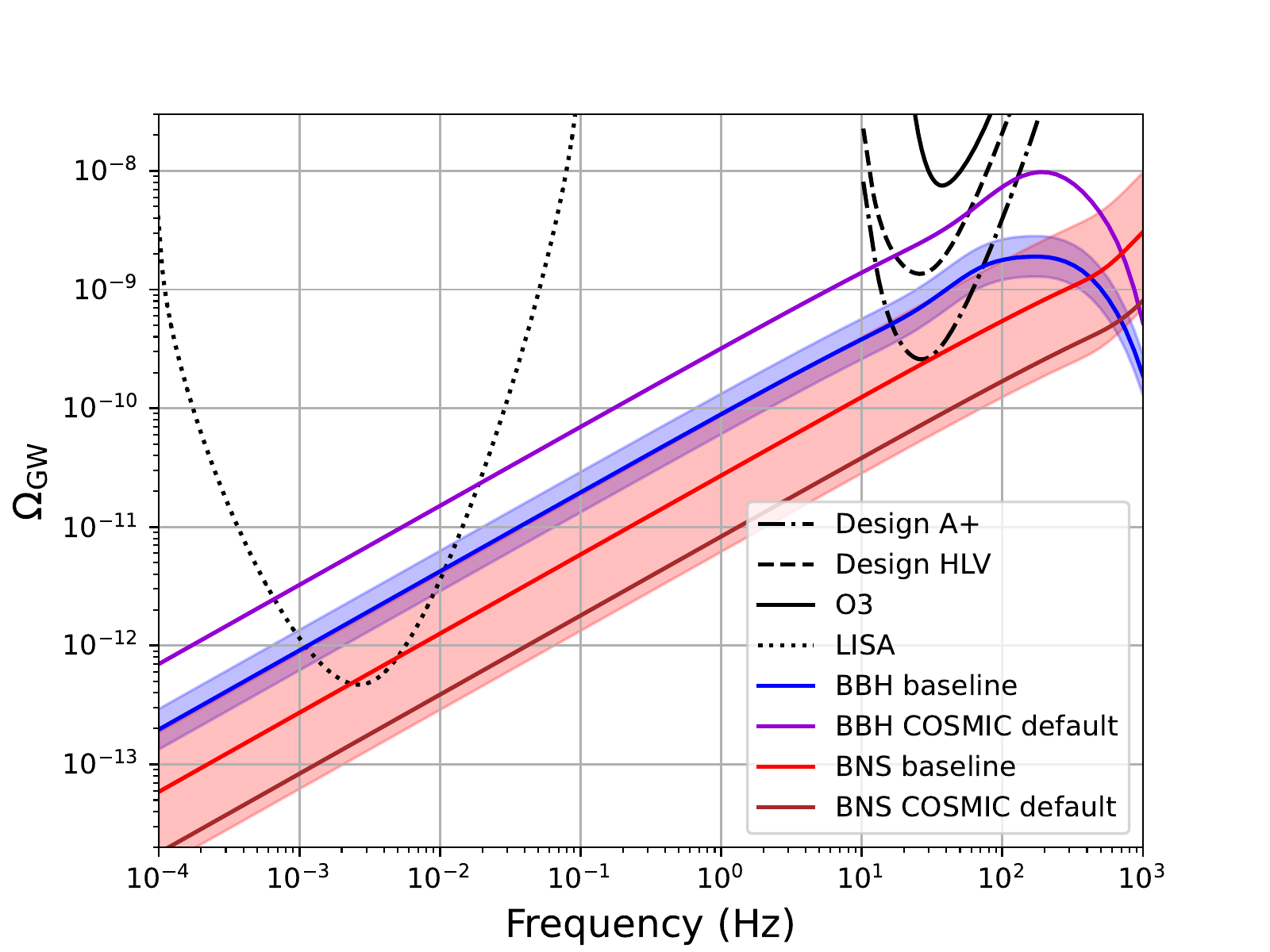}
\caption{The SGWB from both BBHs and BNSs in the frequency range of LISA and LIGO/Virgo. The analytical and population synthesis models for the BNS background are compatible, within the error bars of the former, contrary to their predictions for the BBH.}
\label{fig:summary}
\end{figure}

\begin{table}
\begin{tabular}{|c|cc|cc|}
\cline{2-5}
                                                & \multicolumn{2}{c|}{BBH}          & \multicolumn{2}{c|}{BNS}          \\ \hline
\multicolumn{1}{|c|}{Models}                    & \multicolumn{1}{c|}{\textsc{baseline}} & \textsc{COSMIC} & \multicolumn{1}{c|}{\textsc{baseline}} & \textsc{COSMIC} \\ \hline
\multicolumn{1}{|c|}{$(25\,\mathrm{Hz})\,\,\omgw\cdot10^{10}$} & \multicolumn{1}{c|}{6.83}     &    24.11    & \multicolumn{1}{c|}{2.07}     &    1.60    \\ \hline
\multicolumn{1}{|c|}{$(3\,\mathrm{mHz})\,\,\omgw\cdot10^{12}$} & \multicolumn{1}{c|}{1.89}     &    6.75   & \multicolumn{1}{c|}{0.52}     &    0.40    \\ \hline
\end{tabular}
\label{tab:sgwb_ampl}
\caption{Our predictions for the SGWB in LIGO/Virgo and LISA frequency bands, comparing the \textsc{baseline} and default COSMIC models.}
\end{table}

In Figure \ref{fig:summary} and Table \ref{tab:sgwb_ampl} we compare the backgrounds due to BBH and BNS mergers for the \textsc{baseline} and the \textsc{default} COSMIC models. The uncertainty bands for the \textsc{baseline} model are based on the 90\% credibility interval of the local merger rate measurement by LIGO/Virgo, as reported in \citet{the_ligo_scientific_collaboration_population_2022}, and are consistent with the recent results reported in \citet{babak2023stochastic}. Our results show that the default parameters of COSMIC are not in agreement with the simple analytical model for the BBH background (this is not so surprising since the merger rates predicted by these models are perceptibly different, see Fig.~\ref{fig:Rmerg}). However, this is not the case for the BNS background. It is important to note that the uncertainty  for the BNS case is very high given that LIGO/Virgo have only detected two BNS mergers to date. The fourth run of LIGO/Virgo is expected to detect a few additional BNS mergers \citep{colombo_multi-messenger_2022}, which will result in a more constrained local merger rate. Finally, our results suggest that the BNS background is likely subdominant to the BBH contribution, consistent with previous studies.

\section{Conclusion}\label{sec:conclusion}

We investigated the SGWB produced by several population models of BBHs and BNSs in the frequency ranges of LIGO/Virgo and LISA. We developed three analytical models, namely \textsc{baseline}, \textsc{baseline\_delays}, and \textsc{metallicity cut}, and complemented them with a set of population synthesis models based on the COSMIC code. Our \textsc{baseline} model assumes a merger rate that follows the star formation rate with zero delay times, while the \textsc{baseline\_delays} model takes into account the time delay between the formation of the stellar progenitors and the merger of the CBs. The \textsc{metallicity cut} model includes also the effect of metallicity on the efficiency of producing CBs.

We specifically focused on the mass distribution of CBs in our models. For BBHs, we used mainly the \textsc{Powerlaw+peak} mass distribution from GWTC-3 and investigated some other distributions from the same catalog. For BNSs, we used the mass distribution obtained from Galactic observations, assumed to be valid at all redshifts.

To complement our analytical models, we investigated three models for the BBH population based on the population synthesis code COSMIC, which differ by the set of parameters used to describe the stellar physics, namely \textsc{optimistic}, \textsc{pessimistic}, and \textsc{default}. The mass distribution predicted by the COSMIC models differs from the \textsc{Powerlaw+peak} model, but the time delay distribution was consistent with a simple power law as used in our analytical description. For BNSs, we used only one COSMIC simulation with the \textsc{default} set of parameters.

For BBHs, our analytical models predict $\omgw$ in the range $[4.10^{-10}-~1.10^{-9}]$ at 25~Hz) and $[1.10^{-12}-~4.10^{-12}]$ at 3~mHz, where the range of our predicted values corresponds to the uncertainty in the physical assumptions of our models. These backgrounds could be detectable by LISA with a period of observation of 4 years, but they are unlikely to be detected during the upcoming LIGO/Virgo/Kagra O4 run. However, the O4 run can help to constrain the parameters of our models.

Our analytical models are calibrated to the observed merger rate at $z=0.2$. Thus, the uncertainty in this measurement can lead to a possible variation of about a factor of $1.5$ for BBHs and $2$ for BNSs in the predicted background.

We also investigated the impact of the mass distribution of BBHs on the background, which could vary by a factor of $2$ by varying the mass distribution model within the confidence limits of the GWTC-3 population analysis. Additionally, we found that the uncertainties of the Gaussian peak of the PL+P mass distribution are dominated by the uncertainty in the amplitude of this peak and could lead to a factor of $1.5$ variation in the SGWB. We also discussed the possibility that the position of this peak depends on redshift, but more constraints on the amplitude are needed in order to explore this potential effect. Indeed, our results suggest that the main source of uncertainty is the amplitude of the peak, while its redshift dependence has a negligible impact on the amplitude of the SGWB.

The SGWB predicted by our three COSMIC models varies between $[2.10^{-10}-~2.10^{-9}]$ (25~Hz) and $[7.10^{-13}-~7.10^{-12}]$ (3~mHz). This range, which is larger than the uncertainty due to the measurement error of the local merger rate, corresponds to the uncertainties in the physics of massive stellar binaries \citep{Srinivasan+23}. The \textsc{optimistic} COSMIC model appears to be excluded by observations, as no background has been detected during O3. However, the upcoming O4 run will likely help us constrain the parameters of the stellar model.

Finally, all of our models predict an O$(10)$ number of BBHs and no BNSs that could be individually detectable by LISA for a period of observation of $4$ years.

While we have explored some uncertainties in the SGWB from CBs, several important effects were not included and could lead to further variations in the predicted SGWB. Firstly, we assumed that all binaries are circularized, however including eccentricity and precession in the waveforms might have an important effect on the SGWB amplitude \citep{zhao_stochastic_2020}.

More importantly, in this study we considered only the isolated binary formation scenario for BBHs, while the properties of the observed BBH population suggest that some sources could be formed through the dynamical channel, in particular hierarchical mergers in dense stellar environments. The mass and redshift distribution of this population are expected to be quite different and lead to a different component of the SGWB \citep{perigois_gravitational_2022}. The remnants of Pop III stars, that could have higher merger rates at higher redshifts, could also have an important contribution to the SGWB \citep{dvorkin_metallicity-constrained_2016,2016MNRAS.460L..74H,2022ApJ...940...29M}. These contributions and their associated uncertainties will be studied in future work.

\section*{Acknowledgments}\label{sec:thanks}

This work was supported by the Programme National des Hautes Energies of
CNRS/INSU with INP and IN2P3, co-funded by CEA and CNES. We thank Cyril Pitrou and Jean-Philippe Uzan for their very useful comments and all the interesting discussions we had together.
Clément Pellouin acknowledges funding support from the Initiative Physique des Infinis (IPI), a research training program of the Idex SUPER at Sorbonne Universit\'e.
Rahul Srinivasan and Astrid Lamberts acknowledge support from the graduate and research school EUR SPECTRUM.
This work made use of the Infinity computing cluster at IAP.

\section*{Data Availability}\label{sec:DA}

The data underlying this article will be shared on reasonable request to the corresponding author.\\

\appendix
\section*{Appendix}\label{sec:appendix}

The energy loss of a binary system by gravitational radiation is expressed according to the three phases of the coalescence, the inspiralling  phase (for $f_s < f_\mathrm{merg}$) the coalescence phase (for $f_\mathrm{merg} \leq f_s < f_\mathrm{ring}$) and finally the ringdown phase, i.e., the relaxation phase (for $f_\mathrm{ring} \leq f_s < f_\mathrm{cut}$). It is therefore expressed as  \citep{ajith_template_2009}:
\be
\nonumber
\frac{\mathrm{d}E_{GW}(f_s)}{\mathrm{d}f_s} =
\ee
\\
$$
\frac{\mathrm{d}E_{GW}^\mathrm{Newton}(f_s)}{\mathrm{d}f_s} \times \left\{
    \begin{array}{lll}
        (1+\Sigma_2^3 \alpha_i \nu^i)^2 & \text{for } f_s < f_\mathrm{merg}  \\
        f_s w_m (1+\Sigma_1^2 \epsilon_i \nu^i)^2 & \text{for } f_\mathrm{merg} \leq f_s < f_\mathrm{ring}\\
        f_s^{1/3}w_r\mathcal{L}^2(f_s,f_\mathrm{ring},\sigma) & \text{for } f_\mathrm{ring} \leq f_s < f_\mathrm{cut}
    \end{array}
\right.
$$

with:
    \begin{itemize}
         
        \item $\frac{dE_{GW}^\mathrm{Newton}(f_s)}{df_s} = \frac{5}{12} (G\pi)^{2/3}\mathcal{M}_c^{5/3}F_\theta f_s^{-1/3}$\\
        \item $F_\theta = \frac{(1+\cos^2{\theta})^2}{4} + \cos^2{\theta}$\\
        \item $\nu = \frac{(\pi G M f_s)^{1/3}}{c}$\\
        \item $\mathcal{L}(f,f_\mathrm{ring},\sigma) = \frac{\sigma}{2\pi[(f-f_\mathrm{ring})^2 + 0.24\sigma^2]},\, w_m, \,w_r$ are the continuity coefficients.
    \end{itemize}

The coefficients $\alpha_i$ and $\epsilon_i$ are given in \citet{ajith_template_2009}, with our null assumption for the spin ($\chi = 0$), they can be written as $\epsilon_1 = -1.8897$, $\epsilon_2 = 1.6557$, $\alpha_2 = -\frac{323}{224} + \frac{451}{168}\eta$, $\alpha_3 = 0$, with $\eta = \frac{M_1M_2}{M^2}$.

The phase transition frequencies $f_\mathrm{merg}$, $f_\mathrm{ring}$, $f_\mathrm{cut}$ and $\sigma$ are calculated as \citep{ajith_template_2009}:
\be
\nonumber
\frac{\pi G M}{c^3} \mu_k = \mu_k^0 \text{ with } \mu_k = (f_\mathrm{merg}, \,f_\mathrm{ring},\, f_\mathrm{cut}, \,\sigma)
\ee
where the coefficients $\mu_k^0$ are given in table I of \citet{ajith_inspiral-merger-ringdown_2011}, for $\chi = 0$ we get:
\begin{center}
\begin{tabular}{|l|l|}
   \hline
   $\mu_k$ & $\mu_k^0$\\
   \hline
   $f_\mathrm{merg}$ & $0.066$ \\
   $f_\mathrm{ring}$ & $0.185$ \\
   $f_\mathrm{cut}$ & $0.3236$ \\
   $\sigma$ & $0.925 $\\
   \hline
\end{tabular}
\end{center}

\bibliographystyle{mnras}
\bibliography{biblio}

\end{document}